\documentclass[aps,preprint,longbibliography]{revtex4-1}%
\usepackage{amsfonts}
\usepackage{amsmath}
\usepackage{amssymb}
\usepackage{graphicx}%
\providecommand{\U}[1]{\protect\rule{.1in}{.1in}}
\begin{document}

\title{Velocity jump in the crack propagation induced on a semi-crystalline polymer
sheet by constant-speed stretching}
\author{Takako Tomizawa and Ko Okumura\\Department of Physics, Ochanomizu University, 2-1-1, Otsuka, Bunkyo-ku, Tokyo,
112-8610, Japan}

\begin{abstract}
It has long been known for elastomers that the velocity of crack propagation
jumps as a function of strain. On the other hand, such a jump has not been
reported in the literature for polymers which do not exhibit a rubbery plateau
in the storage-modulus plot. Here, we report observation of jumps in crack
propagation for semi-crystalline polymer sheets without the rubbery plateau,
as a result of pulling the sheets at a constant speed. We discuss the
advantages of this crack-propagation test under constant-speed stretching and
provide physical interpretation of the velocity jump observed for
non-elastomer sheets on the basis of a recently proposed theory for the
velocity jump in crack propagation.

\end{abstract}
\date{\today}
\maketitle

\section{Introduction}

Toughening of polymer materials is an important current problem. Tough
polymers could even replace glass and metal used in automobiles to reduce
their weight, which should greatly contribute to the reduction of energy
consumption. For developing tough polymers, one of the relevant issues is the
velocity-dependent properties of fracture and, in fact, this issue has been
explored for various forms of polymers, which include adhesive
\cite{GentPeelingRate1972,Chaudhury1999Rate,morishita2008contact,Creton2002Adhesion,bhuyan2013crack}%
, laminar \cite{Kinloch1994peelingRate} and viscoelastic polymers
\cite{GreenwoodJohnsonRate,schapery1975theory}, weakly cross-linked gels
\cite{PGGtrumpet,saulnier2004adhesion}, biopolymer gels \cite{lefranc2014mode}%
, and biological composites \cite{bouchbinder2011viscoelastic}.

In the rubber industry, the so-called pure shear test has long been performed
for sheet samples of rubbers to estimate their toughness
\cite{lake2003fracture}. This test is performed in the following manner under
a static boundary condition: (1) Firstly, a sheet is given a fixed strain
through the top and bottom grips. (2) Secondly, the sheet is cut at one of the
free side-edges with keeping the distance between the grips fixed. (3)
Thirdly, the propagation speed of the crack is measured when it reaches a
constant. (The constant speed is attained when the crack propagation length is
long enough compared with the distance between the grips.) When this
measurement is repeated numerous times at different given strains, the
crack-propagation velocity is given as a function of the strain (or the energy
release rate, which is identical to the fracture energy in this case).

It has long been known that the velocity of crack-propagation jumps as a
function of the strain and that the jump is used to control toughness. This
velocity jump observed in elastomers has actively been studied experimentally
\cite{Thomas1981RubberFrac,Tsunoda2000,MoridhitaUrayama2016PRE,morishita2017crack}
and numerically \cite{kubo2017velocity}. Recently, an analytical model has
been proposed to clarify the physical origin of the velocity jump
\cite{sakumichi2017exactly}, and the model is semi-quantitatively compared
with experiments \cite{Okumura2018}.

Although the velocity jump in the crack propagation under the static boundary
condition is useful for evaluating toughness in developing new polymer-based
materials in a sense that such a dangerous jump is better to be avoided for
tough materials, the static crack-propagation test has been limited due to two
factors: (1) No previous studies reported such jumps for resins other than
elastomers. (2) Test requires numerous numbers of sample sheets and repetition
of crack-propagation experiments to determine the velocity at jump.

In this study, we report the velocity jump for sheets of non-elastomer
semi-crystalline polymers by performing crack-propagation test under a dynamic
boundary condition, i.e., under constant-speed stretching, in which case the
velocity at jump is determined by performing a single crack-propagation test.
We demonstrate that this type of crack propagation test is much more sensitive
to detect the jump and discuss the physical interpretation of the jump on the
basis of the analytical theory \cite{sakumichi2017exactly}. Our study paves
the way for a wide use of the velocity jump in developing tough polymer-based materials.

\section{Results and discussions}

\subsection{Materials}

\begin{figure}[h]
\begin{center}
\includegraphics[width=\textwidth]{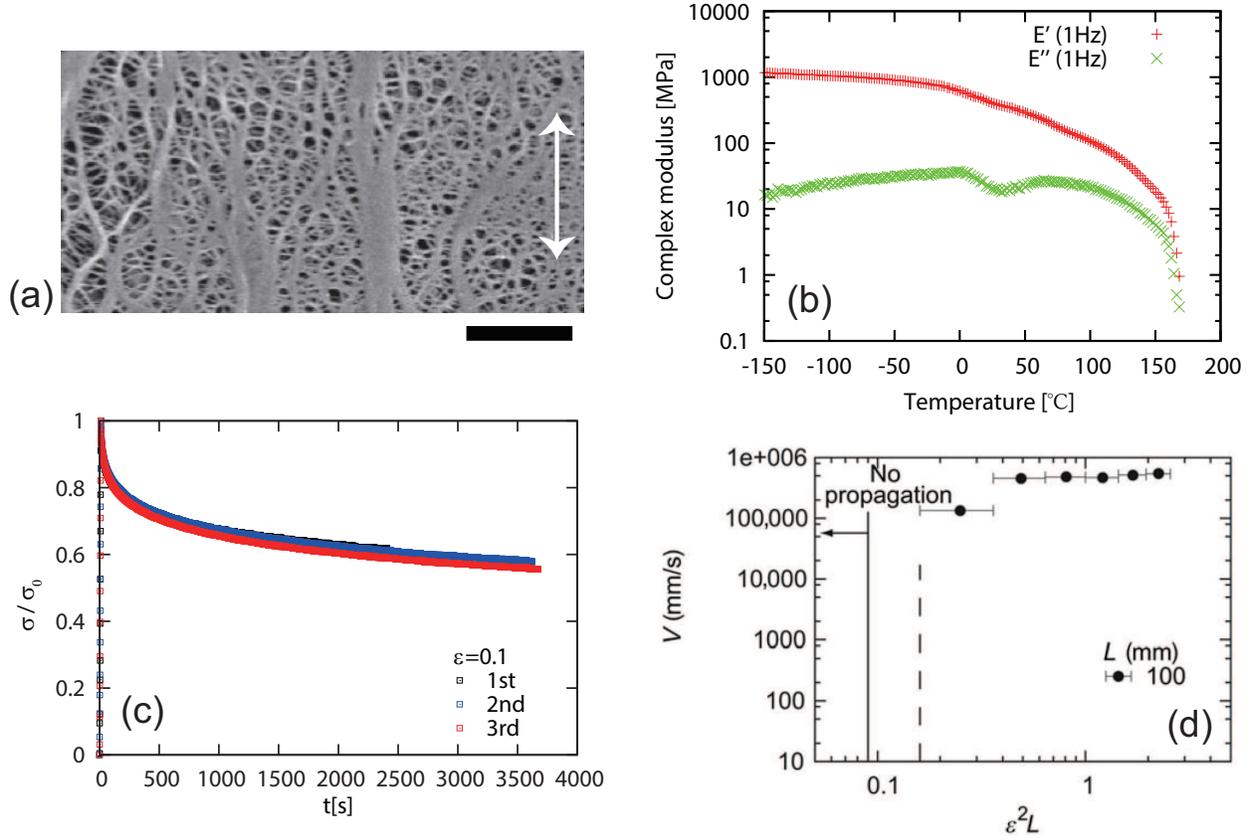}
\end{center}
\caption{(a) A SEM image of the porous polypropylene sheet used in the
experiment. (b) Complex moduli as a function of temperature obtained at 1 Hz.
(c) Stress relaxation curve obtained at the fixed strain of 0.1 for a sample
of width 500 mm and of height $L=$ 100 mm. The stress is renormalized by the
maximum stress $\sigma_{0}$. (d) Crack-propagation velocity vs. a measure of
energy release rate $\varepsilon^{2}L$ under a fixed strain $\varepsilon$
reported in the previous study \cite{Takei2018}. (a) and (b): Copyright 2018
by Mitsubishi Chemical Corporation. (d) Reprinted from \cite{Takei2018} (CC BY
4.0).}%
\label{Fig1}%
\end{figure}

In this study, we used sheets of a porous polypropylene provided by Mitsubishi
Chemical Corporation. The thickness is $h=23$ $\mu$m, the typical size of pore
is of the order of submicrons to several microns (see Fig. \ref{Fig1} (a)),
and the volume fraction is 0.44. The real and imaginary parts of the complex
modulus are shown in Fig. \ref{Fig1} (b). The viscoelastic measurement is
performed for a sample of width 4 mm and length 35 mm by giving a
pre-stretching force of 0.2 N under a strain oscillation of amplitude 0.1 \%
with frequency 1 Hz. The melting point obtained by DSC performed under the
temperature-increase rate 2%
%TCIMACRO{\U{2103}}%
%BeginExpansion
${}^{\circ}{\rm C}$%
%EndExpansion
/min. in the range from 30 to 200%
%TCIMACRO{\U{2103} }%
%BeginExpansion
${}^{\circ}{\rm C}$
%EndExpansion
is 172
%TCIMACRO{\U{2103} }%
%BeginExpansion
${}^{\circ}{\rm C}$
%EndExpansion
(The endothermic peak corresponding to the melting of crystalline domains is
relatively sharp). The crystalline degree of the polypropylene is
approximately 54 per cent. This estimation is based on the melting enthalpy of
the porous film $\Delta H$ = 112 J/g and the equilibrium melting enthalpy of
bulk polypropylene $\Delta H_{0}$ = 209 J/g ($\Delta H/\Delta H_{0}$ = 0.54).
We show the stress relaxation curve in Fig. 1 (c), which suggests that the
stress relaxes significantly within 30 minutes (see Sec. II D). In Fig.
\ref{Fig1} (d), a result of the previous study \cite{Takei2018} is shown for
later convenience.

\subsection{Stress-strain relation}

\begin{figure}[h]
\begin{center}
\includegraphics[width=\textwidth]{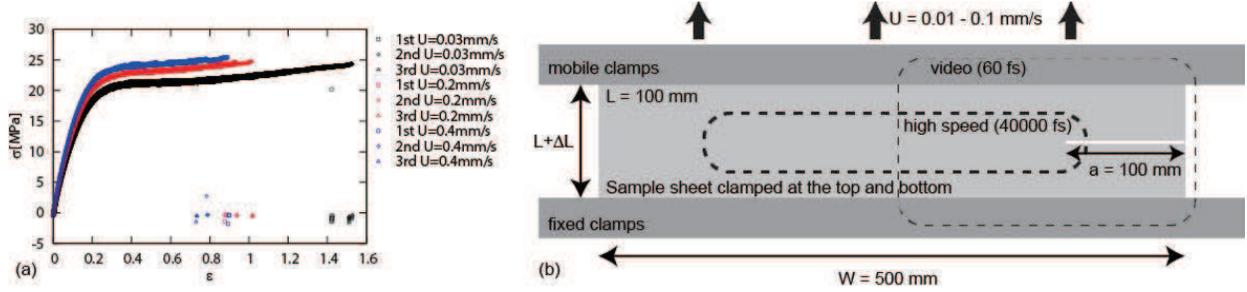}
\end{center}
\caption{(a) Stress vs. strain at three different pulling velocities. (b)
Schematic illustration of the velocity jump experiment, where the areas
covered by two cameras with different frame rates are indicated.}%
\label{Fig2}%
\end{figure}

The stress-strain relation shown in Fig. \ref{Fig2} (a) was measured at three
different pulling speeds in the range from $U=0.02$ mm/s to 0.4 mm/s. The
measurements were performed for the sheet samples of width $W=50$ mm and
height $L=125$ mm by clamping the top and bottom edge of the sample in a
configuration similar to the one shown in Fig. \ref{Fig2} (b) but without the
initial crack of length $a$. (See the text below for further details.) For
convenience, the stress is here defined as the force divided by the (initial)
cross-section of the sheet sample $Wh$.

At each velocity $U$, three measurements were performed, which are well
superposed with each other. However, the breaking point indicated by the
symbol marks that abruptly drop from the smooth curves is weakly dependent on
sample inhomogeneity. The breaking strain tends to be larger when $U~$is
small. The yielding stress is around 20-25 MPa and decreases with $U$, while
the yielding stress around $\varepsilon=0.2$ is almost $U$-independent. These
trends are intuitively natural, because as $U$ becomes small polymer chains
have longer time for relaxation.

\subsection{Crack-propagation test under a constant-speed stretching}

The experiment is schematically explained in Fig. \ref{Fig2} (b). We clamped
the top and bottom edges of a sample sheet of width $W=500$ mm and of initial
height $L=100$ mm and introduced an initial crack of length $a=100$ mm at one
of the side edges. We then observed crack propagation from the initially
introduced crack tip to the other side edge with pulling the top clamps at a
fixed speed $U$ in the range from 0.01 - 0.1 mm/s. The movement of the clamps
is controlled by a slider system (EZSM6D040 K, Oriental Motor). The quantity
$\Delta L$ is the increase in the height at each moment after stretching starts.

\begin{figure}[h]
\begin{center}
\includegraphics[width=\textwidth]{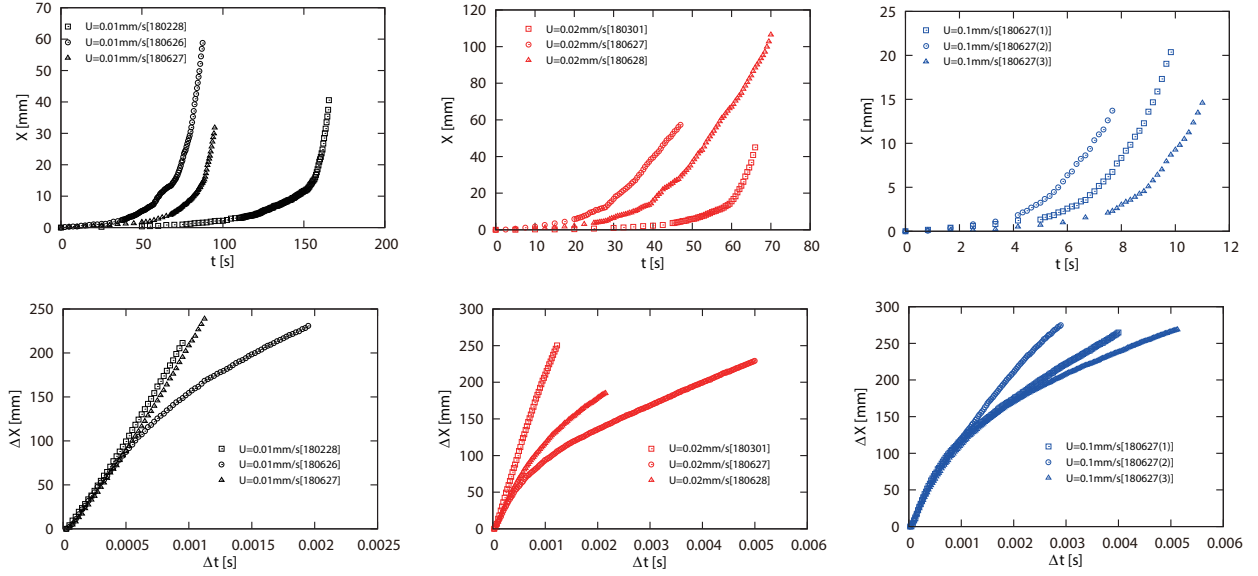}
\end{center}
\caption{Crack-tip position as a function of elapsed time at three different
stretching velocities $U$. (Top) Results obtained from a 60-fps camera before
the velocity jump. The position $X=0$ and the time $t=0$ correspond the tip
position of the initial crack and the starting time of stretching,
respectively. (Bottom) Results obtained from a 40000-fps camera after the
velocity jump. The position $\Delta X=0$ and the time $\Delta t=0$ correspond
the crack-tip position and the moment at the jump, respectively.}%
\label{Fig3}%
\end{figure}

As a result, we found that the crack-propagation velocity jumps dramatically
at a certain point. To analyze the dynamics before and after the jump, we set
two cameras with significantly different frame rates. To capture the dynamics
before the jump, we used a digital camera with a video function (D800E, Nikon)
and acquired snapshots with the rate of 60 fps. For the dynamics after the
jump, we used a high-speed camera (FASTCAM Mini UX100, Photoron) with a lens
(AF-S NIKKOR 20mm 1:1.8G ED, Nikon). The areas covered by the two cameras are
indicated in Fig. \ref{Fig2} (b). The snapshot obtained with the 60-fps camera
just after the jump records the moment after the crack reaches the opposite
sample end. On the contrary, before the jump, no movements of the crack tip
are recorded by the 40000-fps camera because of the limited memory storage of
the high-speed camera.

In the following, we explain the results shown in Figure 3. The data shown in
the figure seem to exhibit strong sample dependence. However, considering
experimental artifacts (difficulty in determining the stress-zero state in the
case of Fig. 3(a) and an edge effect in the case of Fig. 3 (b)), sample
dependence is in fact less strong than its appearance, as explained below.
This is reasonable considering that bulk properties characterized in Fig 1 and
Fig 2 by the rheology measurements and the stress-strain curve do not show
strong sample dependence, although, in principle, the local distribution of
pore size could affect the crack propagation.

The three graphs in the top panel of Fig. \ref{Fig3} show the results for the
dynamics before the jump obtained by the 60-fps camera at different stretching
velocities $U$. We see that there is a rather strong sample dependence.
However, the dependence might be enhanced because of the difficulty in
determining the zero-strain state, which corresponds to $t=0$. For example, at
$U=0.1$ mm/s, the three plots are well superposed with each other if we allow
time-shifts of the order of a few seconds, which implies that the sample
dependence is in fact relatively small. This order of time-shifts suggests a
rough estimate for the magnitude of error in the increase in height $\Delta
L$, which is about $0.1$ mm (i.e., $U$ times a few seconds). This corresponds
to the error in the strain $\simeq0.001$ because $L=100$ mm. Similar estimates
for the errors in $\Delta L$ and the strain are obtained from the two
remaining values of $U$. For example, at $U=0.01$ mm, the time-shifts of 10
seconds would decrease the sample dependence significantly (i.e., it would
make the three plots well superposed).

The three graphs shown in the bottom panel of Fig. \ref{Fig3} show the results
for the dynamics after the jump obtained by the 40000-fps camera at different
stretching velocities $U$. Just after the jump, the sample dependence of the
data is fairly small. On the contrary, the speed (the slope) tends to decrease
as the crack tip approaches the side edge opposite to the one at which the
initial crack is introduced. Such a decrease in velocity would not be observed
if the sample is long enough in the direction of crack propagation. This is
called the edge effect and we try to avoid this effect in our velocity
analysis presented below. (Since a slight error in identifying the zero-strain
state leads to a significant difference in the position of the jump, apparent
sample dependence due to the edge effect will be enhanced compared with real
sample dependence.)

\begin{figure}[h]
\begin{center}
\includegraphics[width=\textwidth]{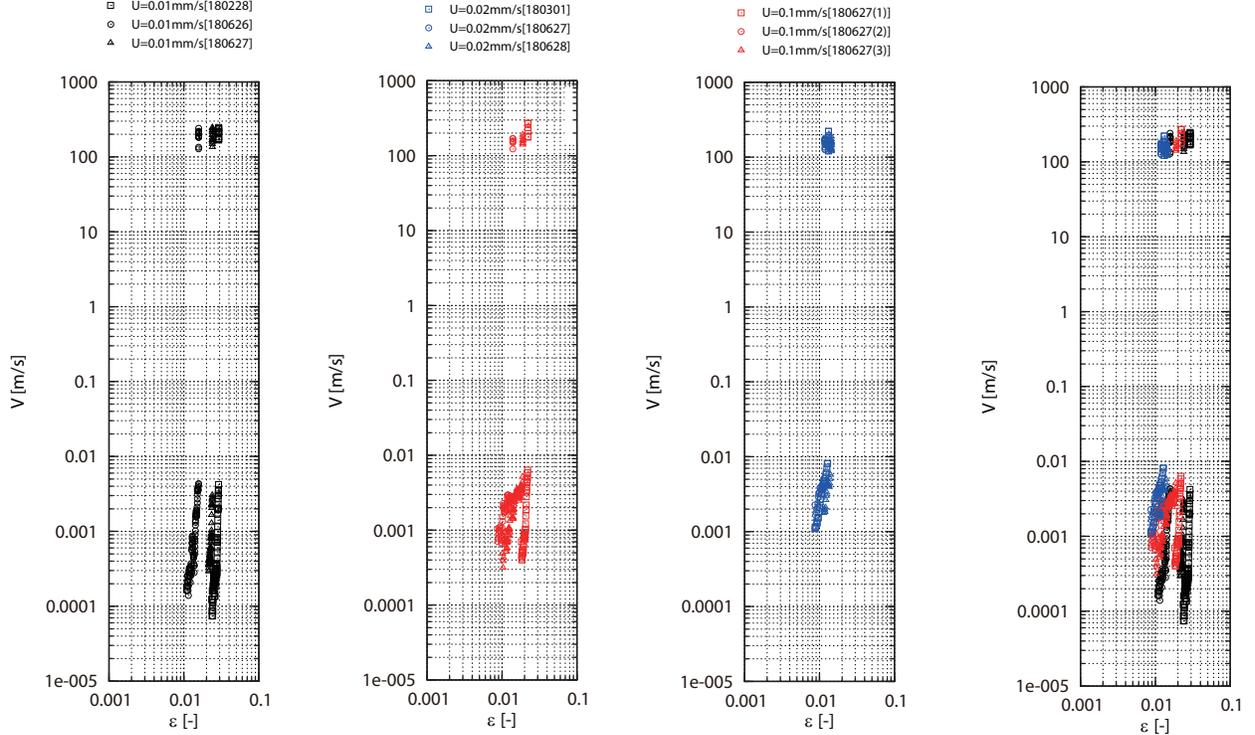}
\end{center}
\caption{Crack-propagation velocity vs strain at three different stretching
velocity.}%
\label{Fig4}%
\end{figure}

The crack-propagation velocity obtained from the data shown in Fig. \ref{Fig3}
is given in Fig. \ref{Fig4}. The velocity before the jump is obtained from the
data shown in the top panel of Fig. \ref{Fig3}. (The $i$th velocity $V_{i}$ is
obtained as the backward difference $V_{i}=(X_{i}-X_{i-1})/\delta t$ (with
$\delta t=1/60$ sec) from the smoothed $i$th position $X_{i}$ given as
$(x_{i}+x_{i-1}+\cdots+x_{i-4})/5$ where $x_{j}$ is the $j$th position of the
raw data.) The velocity after the jump is obtained from the data shown in the
bottom panel of Fig. \ref{Fig3} by using the initial region in which the
sample dependence is small to avoid the edge effect. (The velocity is obtained
as the forward difference $V_{i}=(x_{i}-x_{i+1})/\delta t$ (with $\delta
t=1/40000$ sec) of the raw $\Delta X-\Delta t$ data in the ranges $\Delta t=0$
to 0.4 ms and $\Delta t=0$ to 0.2 ms for $U=0.01$ and 0.1 m/s and for $U=0.02$
m/s, respectively.)

Although there exists the sample dependence probably originating from
inhomogeneous porous structures, which tends to be enhanced because of
experimental artifacts, some robust features are visible in Fig. \ref{Fig4}.
(1) The velocity changes nearly four orders of magnitude at the jump. (2)
Before the jump the crack-propagation velocity is in the range around from
0.0001 to 0.01 m/s, which corresponds to the strain-rate range from 0.001 to
0.1 s$^{-1}$. (3) After the jump, the velocity reaches around a few hundreds
m/s. (4) The strain at the jump tends to decrease as the pulling velocity $U$
increases. (5) Before the jump, the crack-propagating velocity $V$ as a
function of $\varepsilon$ is significantly large for high $U$. (6) After the
jump, $V$ tends to be small for high $U$ and $V$ seems to be on a straight
line as a function of $\varepsilon$ on the log-log plot. (7) The size of the
jump tends to decrease as $U$ increases.

The strain range for the data in Fig. 4 is around 0.01 to 0.04. In the
corresponding strain range, the stress-strain relation is practically linear
and free from the effect of plastic flow as seen in Fig. 2(a). This implies
that the strain $\varepsilon$ and fracture energy $G$ (i.e., energy release
rate) satisfy a simple linear relation on a log-log scale ($\log G$ = const.+
$2\log\varepsilon$). Accordingly, we here used the strain $\varepsilon$
(instead of fracture energy $G$) for the horizontal axis.

\subsection{Crack propagation test under a fixed-grip condition}

We now discuss the previous results shown in Fig. \ref{Fig1} (c)
\cite{Takei2018}, which is obtained by observing a constant-velocity crack
propagation under a fixed strain. By repeating crack-propagation experiments
with changing the value of the fixed strain, one obtains the crack-propagation
velocity as a function of the given strain. (We need at least one sample sheet
to obtain a single point on the $V-\varepsilon$ plot.) Here, the velocity is
given as a quantity $\varepsilon^{2}L$, where $L$ is the height of the sample.
This quantity is proportional to the so-called energy release rate and the
velocity is expected to be a universal function of this quantity for a certain
material although $L$ are different. The experimental geometry is similar\ to
the one given in Fig. \ref{Fig2} (b). The differences are that the mobile pair
of clamp is fixed to give a certain value of $\Delta L$ and that the small
initiating crack is introduced at one of the side edges after $\Delta L$ is
given. We note here for later convenience that this process of preparation,
i.e., giving a fixed strain to the sheet before introducing a small cut to
initiate crack propagation, requires a finite amount of time.

In this previous experiment, we failed to observe the velocity jump. Instead,
we found that the crack propagates nearly at a constant speed around $500$ m/s
for $\varepsilon$ larger than around 0.07 ($\varepsilon^{2}L\simeq0.5$) and
that the crack does not propagate at all for $\varepsilon~$smaller than around
0.03 ($\varepsilon^{2}L\simeq0.1$)$.$

\subsection{Comparison of the results under the dynamic and static boundary
conditions}

We find clearly that the velocity jump is observed in the present experiment
performed under the dynamic boundary condition, whereas the jump is not
observed in the previous experiment performed under the static condition. The
high-velocity regime after the jump in the present study is fairly consistent
with the high-speed crack propagation observed in the previous study. (In the
present study the high-velocity regime after the jump is "short," i.e.,
observed in a small strain range at a given $U$. This does not suggest
difference but is merely because the sample length is not enough: after the
jump, the crack soon reaches the opposite side end of the sample.) However,
the low-velocity regime before the jump in the present study is absent in the
previous study, which marks the significant difference between the two experiments.

We consider the principal reason for not observing this regime under the
static boundary condition is that the stress relaxation occurred in the
above-mentioned preparation time for the fixed-grip experiment. In the
previous experiment, the preparation time was of the order of 30 minutes,
during which the stress is significantly relaxed (because this material is a
linear viscoelastic material for small strains) as suggested in Fig. 1 (c) to
suppress the crack propagation. On the contrary, when the given strain is
relatively high, because this material exhibits yielding behavior (see Fig.
\ref{Fig2} (a)), the induced plastic deformation tends to suppress relaxation,
and thus the crack-propagation behavior under the static condition becomes
similar to that under the dynamic condition.

Note that under the dynamic boundary condition in the present study, there is
no preparation time and the effect of stress relaxation is practically
suppressed. By definition, the preparation time is the time duration from the
moment one starts to give a fixed strain to the sample to the moment one
introduces a cut at one of the free side edges after one finished giving the
fixed strain. In the experiment under the dynamic boundary condition, crack
keeps propagating with the given strain simultaneously increasing. This means
that the dynamic test is free from preparation time and the effect of stress
relaxation (during the preparation time) is by definition absent in the
dynamic test.

\subsection{Plausible physical pictures emerging from the previous theory}

In our previous study \cite{sakumichi2017exactly}, we showed that the glass
transition that occurs near the very localized vicinity of the crack tip can
trigger the velocity jump. This theory suggests that the velocity jump could
be observed if the storage modulus possesses the glassy regime in the high
frequency range and the rubbery regime in the low frequency range. The
high-velocity regime after the jump reflects the glassy dynamics, whereas the
low-velocity regime before the jump corresponds to the rubbery dynamics.

We consider the velocity jump observed in the present study is understood on
the basis of this previous theory. As shown in Fig. \ref{Fig1} (c), this
material clearly has the glassy regime and shows the glass transition
temperature around at 0 $^{\circ}$C. But, it starts to melt around at 170
$^{\circ}$C, showing no rubbery plateau (or before showing the rubbery
plateau). This implies that we cannot access the rubbery regime of this
material by raising temperature. However, on the basis of the time-temperature
correspondence, we may access the rubbery regime by increasing the time scale
without facing melting of the sample, which we consider is realized by the
present crack-propagation experiment. Because of the slow dynamics induced by
the slow crack propagation, the material near the crack starts to behave like
a rubber. This is our physical interpretation of the slow-velocity regime
before the velocity jump. In other words, the present result is consistent
with and even expected from the previous theory, if we assume that our
previous theory is applicable to the present case.

\subsection{Advantages of the dynamic crack-propagation test}

We here discuss the advantages of the dynamic crack-propagation test. As seen
above, the dynamic test could be much sensitive to the velocity jump because
it is less subject to the effect of strain relaxation. In addition, the
velocity at the jump at each pulling velocity $U$ is precisely detected
because the strain value is dynamically and continuously changed with time. In
other words, a fine tuning of strain is possible to obtain the precise value
of strain at the jump. An important practical advantage is that the dynamic
test is much more timesaving and needs much less amount of samples. This is
because in the case of the static test we need one sample to obtain a single
point on the velocity-strain plot. On the contrary, one entire curve for the
velocity-strain relation is obtained from a single sample sheet. Note,
however, that, as already mentioned, the dynamic test may not be practical if
one would like to obtain the high-velocity regime over a wide range: we would
need to have nearly an infinite sample width $W$ in order to obtain a wide
rage of the high velocity range in the dynamic test.

\subsection{Comparison of our results with results obtained from rubbers}

We here compare our results with results obtained from rubbers. In the case of
rubber, the velocity jump occurs typically from $10^{-1}$ mm/s to 1 m/s at a
critical energy release rate. The ratio of the velocity just after the jump
$V_{a}$ to the velocity just before the jump $V_{b}$ is about $10^{4}$, which
is comparable to the value obtained the present case. As for $V_{a}$ and
$V_{b}$, both are about 100 times as large as those of rubbers. According to
the previous theory \cite{sakumichi2017exactly}, $V_{a}$ and $V_{b}$ both
scale with the factor $l/\tau_{R}$, i.e., the length scale $l$ below which the
continuum description breaks down divided by the characteristic time $\tau
_{R}$ for the rubbery regime (see Eq. (16) and (17) in \cite{Okumura2018}).
Compared with rubbers, $\tau_{R}$ tends to be larger because the plots of the
complex elastic moduli tend to be shifted in the high temperature side in the
present case. However, the length $l$ may be much larger in the present case
because of the porous structure, which may explain at least qualitatively why
$V_{a}$ and $V_{b}$ are shifted to the higher side.

In the slow mode regime, in which the propagation velocity $V$ is smaller than
$V_{b}$, the velocity $V$ seems to scale with some power of the fracture
energy $G$ ($\sim\varepsilon^{2}$ in the present case) as in the case of
rubbers: the plot $V$ vs $G$ (or $\varepsilon$) on a log-log scale is on a
straight line. However, the slope of the straight line is much larger in the
present case. This tendency is qualitatively in accord with the experimentally
known fact (for rubbers) that the slope becomes large with the characteristic
length scale $l$ (i.e., cross-linking distance) \cite{Tsunoda2000}. As
stressed above, $l$ is expected to be much larger in the present case and,
thus, from the experimentally known fact, the slope on a log-log scale in the
slow mode regime is expected to be very large, which is supported by Fig. 4.

As for the fast mode, in which $V$ is larger than $V_{a}$, comparison is
generally difficult. This is because as explained in Sec. II E below, the
crack propagation experiment under the dynamic boundary condition as in the
present case, it is difficult to obtain the data in the fast mode regime in a
wide range because of the limitation of the sample length (after the jump $V$
is extremely fast).

\section{Conclusion}

We observed a jump in the crack-propagation speed as a function of applied
strain for a polymer sheet which is not elastomer, i.e., whose storage modulus
does not exhibit the rubbery plateau. The jump was not observed in the
previous study in which the experiment is done under the static boundary
condition. In the present study, by employing the dynamic boundary condition,
we succeeded in observing the jump. The dynamic test demonstrated in the
present study is promising because of a number of advantages such as the
sensitivity to the velocity jump, timesaving and cost-effective features. If
we assume that the previous theory [19] is applicable to the present results,
plausible physical pictures emerge: (1) The velocity jump in the present study
results from the glass transition that occurs in the vicinity of the crack
tip. (2) The low-velocity regime before the jump may probe into the rubbery
dynamics, which is usually hidden by melting and cannot be accessible by
raising temperature.

\section*{Acknowledgements}

The authors thank Dr. K. Tsunoda (Bridgestone) for discussion on the
crack-propagation test under a constant pulling speed. They are grateful to
Mitsubishi Chemical Corporation for providing sample sheets, the SEM image
[Fig. \ref{Fig1} (a)], and the viscoelastic plots [Fig. \ref{Fig1} (b)],
together with the results concerning the melting temperature and degree of
crystallization of the sample. This work was partly supported by ImPACT
program of Council for Science, Technology and Innovation (Cabinet Office,
Government of Japan).

\section*{References}

\bibliographystyle{unsrt}
\bibliography{C:/Users/okumura/Documents/main/JabRef/granular,C:/Users/okumura/Documents/main/JabRef/fracture,C:/Users/okumura/Documents/main/JabRef/wetting}

\begin{thebibliography}{10}

\bibitem{GentPeelingRate1972}
AN~Gent and J~Schultz.
\newblock Effect of wetting liquids on the strength of adhesion of viscoelastic
  material.
\newblock {\em The Journal of Adhesion}, 3(4):281--294, 1972.

\bibitem{Chaudhury1999Rate}
Manoj~K Chaudhury.
\newblock Rate-dependent fracture at adhesive interface.
\newblock {\em The Journal of Physical Chemistry B}, 103(31):6562--6566, 1999.

\bibitem{morishita2008contact}
Yoshihiro Morishita, Hiroshi Morita, Daisaku Kaneko, and Masao Doi.
\newblock Contact dynamics in the adhesion process between spherical
  polydimethylsiloxane rubber and glass substrate.
\newblock {\em Langmuir}, 24(24):14059--14065, 2008.

\bibitem{Creton2002Adhesion}
Costantino Creton, Edward Kramer, Hugh Brown, and Chung-Yuen Hui.
\newblock Adhesion and fracture of interfaces between immiscible polymers: from
  the molecular to the continuum scal.
\newblock {\em Adv. Polymer Sci.}, 156:53--136, 2002.

\bibitem{bhuyan2013crack}
Satyam Bhuyan, Fran{\c{c}}ois Tanguy, David Martina, Anke Lindner, Matteo
  Ciccotti, and Costantino Creton.
\newblock Crack propagation at the interface between soft adhesives and model
  surfaces studied with a sticky wedge test.
\newblock {\em Soft Matter}, 9(28):6515--6524, 2013.

\bibitem{Kinloch1994peelingRate}
AJ~Kinloch, CC~Lau, and JG~Williams.
\newblock The peeling of flexible laminates.
\newblock {\em International Journal of Fracture}, 66(1):45--70, 1994.

\bibitem{GreenwoodJohnsonRate}
JA~Greenwood and KL~Johnson.
\newblock The mechanics of adhesion of viscoelastic solids.
\newblock {\em Phil. Mag. A}, 43(3):697--711, 1981.

\bibitem{schapery1975theory}
RA~Schapery.
\newblock A theory of crack initiation and growth in viscoelastic media.
\newblock {\em International Journal of Fracture}, 11(1):141--159, 1975.

\bibitem{PGGtrumpet}
PG~de~Gennes.
\newblock {\em C. R. Acad. Sci. Paris}, 307:1949, 1988.

\bibitem{saulnier2004adhesion}
F~Saulnier, T~Ondarcuhu, A~Aradian, and E{\=\i}~Rapha{\"e}l.
\newblock Adhesion between a viscoelastic material and a solid surface.
\newblock {\em Macromolecules}, 37(3):1067--1075, 2004.

\bibitem{lefranc2014mode}
Maxime Lefranc and Elisabeth Bouchaud.
\newblock Mode i fracture of a biopolymer gel: Rate-dependent dissipation and
  large deformations disentangled.
\newblock {\em Extreme Mechanics Letters}, 1:97--103, 2014.

\bibitem{bouchbinder2011viscoelastic}
Eran Bouchbinder and Efim~A Brener.
\newblock Viscoelastic fracture of biological composites.
\newblock {\em Journal of the Mechanics and Physics of Solids},
  59(11):2279--2293, 2011.

\bibitem{lake2003fracture}
GJ~Lake.
\newblock Fracture mechanics and its application to failure in rubber articles.
\newblock {\em Rubber Chem. Tech.}, 76(3):567--591, 2003.

\bibitem{Thomas1981RubberFrac}
A~Kadir and AG~Thomas.
\newblock Tear behavior of rubbers over a wide range of rates.
\newblock {\em Rubber Chem. Tech.}, 54(1):15--23, 1981.

\bibitem{Tsunoda2000}
K~Tsunoda, JJC Busfield, CKL Davies, and AG~Thomas.
\newblock Effect of materials variables on the tear behaviour of a
  non-crystallising elastomer.
\newblock {\em J. Mater. Sci.}, 35(20):5187--5198, 2000.

\bibitem{MoridhitaUrayama2016PRE}
Yoshihiro Morishita, Katsuhiko Tsunoda, and Kenji Urayama.
\newblock Velocity transition in the crack growth dynamics of filled
  elastomers: Contributions of nonlinear viscoelasticity.
\newblock {\em Physical Review E}, 93(4):043001, 2016.

\bibitem{morishita2017crack}
Yoshihiro Morishita, Katsuhiko Tsunoda, and Kenji Urayama.
\newblock Crack-tip shape in the crack-growth rate transition of filled
  elastomers.
\newblock {\em Polymer}, 108:230--241, 2017.

\bibitem{kubo2017velocity}
Atsushi Kubo and Yoshitaka Umeno.
\newblock Velocity mode transition of dynamic crack propagation in
  hyperviscoelastic materials: A continuum model study.
\newblock {\em Scientific Reports}, 7:42305, 2017.

\bibitem{sakumichi2017exactly}
Naoyuki Sakumichi and Ko~Okumura.
\newblock Exactly solvable model for a velocity jump observed in crack
  propagation in viscoelastic solids.
\newblock {\em Scientific Reports}, 7(1):8065, 2017.

\bibitem{Okumura2018}
Ko~Okumura.
\newblock Velocity jumps in crack propagation in elastomers: Relevance of a
  recent model to experiments.
\newblock {\em Journal of the Physical Society of Japan}, 87(12):125003, 2018.

\bibitem{Takei2018}
Atsushi Takei and Ko~Okumura.
\newblock Crack propagation in porous polymer sheets with different pore sizes.
\newblock {\em MRS Communications}, pages 1--6, 2018;
  http://dx.doi.org/10.1557/mrc.2018.222.

\end{thebibliography}

\end{document}